**Collective behavior in the kinetics and equilibrium of solid-state photoreaction**

Ruobing Bai, Ying Shi Teh, Kaushik Bhattacharya*

Division of Engineering and Applied Science, California Institute of Technology, Pasadena, California 91125, USA
* Corresponding author: bhatta@caltech.edu

**Abstract**

There is current interest in developing photoactive materials that deform on illumination. The strategy is to develop new photoactive molecules in solution, and then to incorporate these in the solid-state either by crystallization or by inserting them into polymers. This letter shows that the kinetics and the nature of the photo-induced phase transitions are profoundly different in single molecules (solution) and in the solid state using a lattice spin model. In solution, where the molecules act independently, the photoreaction follows first-order kinetics. However, in the solid state where the photoactive molecules interact with each other and therefore behave collectively during reaction, photoreactions follow the sigmoidal kinetics of nucleation and growth as in a first-order phase transition. Further, we find that the exact nature of the photo-induced strain has a critical effect on the kinetics, equilibrium, and microstructure formation. These predictions agree qualitatively with experimental observations, and provide insights for the development of new photoactive materials.

Some molecules can absorb light of a certain wavelength and change their shape, dissociate, or combine to form new molecules. These photoreactions (photo-isomerization, -dimerization, etc.) enable applications including optical grating [1], image storage [2], surface alignment [3], wettability and permeability modulation [4, 5], and photomechanical actuation [6-12]. Photomechanical actuation is particularly attractive because it can be effected at a distance, different frequencies can be used to actuate different modes or control an object, and corrosion-free lightweight fiber optic cables can deliver significant power over long distances. This has motivated the development of a number of new photoactive solids [10-15]. Still, promising molecules in solution do not always lead to effective photoactive solids, and this motivates the current work.

When embedded in a solid, the molecules cannot react independently, but interact collectively through the solid matrix. Despite the extensive studies on single-molecule photoreaction [16-18], studies on solid-state photoreaction are still based on either simple first-order reaction kinetics, or mean-field approaches such as the Johnson–Mehl–Avrami–Kolmogorov (JMAK) model [19-22] and the Finke–Watzky model [23]. The coarse-grained nature of these models limits their capability to explore the more detailed microstructure inside the material, and thus they do not fully explain various observations, such as the sigmoidal kinetics reported by recent experimental evidence [24-26], and the formation of specific domain patterns in the solid during the reaction [27-29].

In this letter, we address the above challenges in solid-state photoreaction by proposing a driven Ising lattice spin model with both nearest-neighbor and long-range intermolecular interactions where the evolution is driven by light-induced photoreaction in addition to the usual thermal relaxation. The light illumination pumps photon energy into the system while the thermal fluctuation drives a relaxation, and the system eventually reaches equilibrium or the *photo-stationary* state. Similar kinds of external driving have been studied in nonequilibrium friction and shear in driven Ising systems [30-32].

We start by describing the energy landscape of azobenzene (a representative photoactive molecule) at the single-molecule level inferred from first principles calculations with the corresponding reaction paths illustrated in FIG. 1a [16-18]. The electronic-ground-state energy landscape has a double-well structure, with two minima corresponding to the *trans*- and *cis*-isomers, *trans* being lower energy than *cis*. Thermal fluctuations exist between these two states following the usual rate kinetics. Also existing is a metastable electronic excited state that is accessed when the *trans*-isomer absorbs a photon. The molecule relaxes back to either the *trans*- or *cis*-state, with a probability determined by the ground-state energy landscape.

We incorporate the above single-molecule framework into a solid of many photoactive molecules by adopting a 2D periodic square lattice of $n \times n$ elements extended periodically (analogous to spins in an Ising model), each taking a value $S_\alpha$ of +1 (*cis*) or -1 (*trans*) (FIG. 1b).

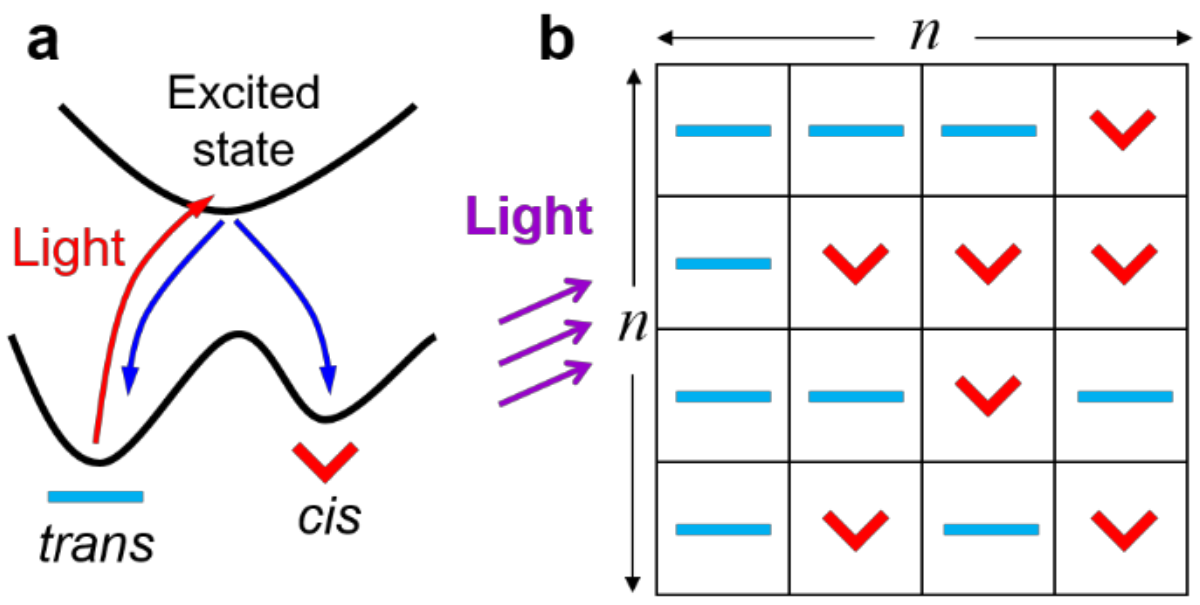

**FIG. 1**. **(a)** Single-molecule energy landscape in the photo-isomerization of azobenzene. **(b)** Ising-like lattice model for photo-isomerization in solid-state azobenzene.

We consider the following Hamiltonian for our Ising model:

$$H = -\frac{1}{2} J \sum_{\langle \alpha,\beta \rangle} S_\alpha S_\beta + h \sum_{\alpha=1}^{n^2} S_\alpha + E\bar{W}\left(S_\alpha\right). \quad (1)$$

The first term represents the nearest-neighbor interaction that describes the weak intermolecular force such as the dispersion force between molecules ($\langle \alpha, \beta \rangle$ denotes all pairs of nearest neighbors, and 1/2 accounts for the interchange of $\alpha$ and $\beta$). We take $J$ to be positive so that neighboring elements prefer the same state. This also gives rise to the interfacial energy between different phases. The second term describes the energetic preference of the *trans*-state with $h$ being positive. The third term describes the long-range elastic interaction. During a solid-state photoreaction, the shape change of molecules may induce non-negligible stress and strain fields in the solid. The *transformation strain* or *photo-strain*, defined as the spontaneous strain of the material under no external stress, in one location affects stress field globally and interacts with a photo-strain at a distal location. Hence, photoreaction induces a long-range elastic interaction [33, 34]. $E\bar{W}\left(S_\alpha\right)$ in (1) is the total elastic energy in the system, where $E$ is the plane strain Young's modulus of the solid and $\bar{W}\left(S_\alpha\right)$ ($\alpha = 1,\ldots,n^2$) is the dimensionless elastic energy that depends on $S_\alpha$ as described later.

The Hamiltonian in (1) describes a variety of materials: (i) a solution, uncrosslinked polymer melt, or dilute-crosslinked polymer network when $J/kT = E/kT = 0$, (ii) moderately crosslinked polymer when $J/kT \neq 0$ and $E/kT = 0$, and (iii) crystals of densely packed photoactive molecules when $J/kT$ small and $E/kT$ large.

To model photo-isomerization using this Hamiltonian, we adopt a Monte Carlo algorithm, where in each step we conduct a single-flip trial following one of two processes. In a *photo activation* process, we flip a random element to the *cis*-state with a probability of min[1, exp($-\Delta H/kT$)] if it is initially *trans*, and do nothing otherwise. This corresponds to the reaction paths at the single-molecule energy landscape illustrated in FIG. 1a. In a *thermal relaxation* process, we follow the standard Metropolis importance sampling by flipping a random element with a probability of min[1, exp($-\Delta H/kT$)] [35, 36]. $\Delta H$ in both cases denotes the change of total Hamiltonian upon flipping. We define the dimensionless light intensity $I$ as the ratio between the photo activation and the thermal relaxation steps. For example, if we conduct photo activation for $N$ steps and thermal relaxation for 1 step, then $I = N$.

With a time scale assigned to each step, the single-flip process of the Monte Carlo simulation can be interpreted kinetically by the Glauber kinetics of the Ising model [36, 37], which allows us to explore the reaction kinetics. We record the evolution of *cis*-ratio with reaction time represented by the increasing MC (Monte Carlo) steps per element (the accumulated number of steps divided by $n^2$).

We conduct our simulations on a 100×100 periodic lattice ($n = 100$), and an initial condition of 0% *cis*-ratio. The total MC steps per element are 2000, large enough for the reaction to reach photo-stationary. The photo-stationary *cis*-ratio is calculated as the average of the last MC step per element ($10^4$ steps). While the results presented here are in two dimensions to manage the computational complexity, we expect to see qualitatively similar results in three dimensions.

We start without the elastic interaction ($E = 0$), and study the difference between photoreactions in solution ($J/kT = 0$) and moderately cross-linked solid ($J/kT > 0$) with fixed $h/kT = 1$ that seeks to drive the material to the *trans*-state. FIG. 2a plots the photo-stationary *cis*-ratio as a function of light intensity $I$. The photo-stationary state in solution ($J/kT = 0$) shows first-order reaction kinetics. When $I = 0$, the photo-stationary *cis*-ratio is finite due to thermal fluctuation since $J/kT = 0$, but is much below 50% due to the positive $h/kT$. The *cis*-ratio increases monotonically with $I$, linearly at small $I$, and approaches 100% exponentially at large $I$. The material remains completely disordered through the reaction as shown in FIG. 2b and Movie 1. The forward reaction rate which has contributions from both the photo activation and the thermal relaxation can be expressed as $\left(1-\phi_{cis}\right)\exp\left(-2h/kT\right)\left(I+1\right)\tau^{-1}$ where $\tau$ is the time constant of each step. The backward reaction rate is governed by thermal relaxation alone and is expressed as $\phi_{cis}\tau^{-1}$. Equilibrating the two rates leads to the photo-stationary *cis*-ratio $\phi_{cis} = 1-\left[1+\left(I+1\right)\exp\left(-2h/kT\right)\right]^{-1}$. This analytical solution is plotted in FIG. 2a as the solid line, showing good agreement with the simulation.

In contrast to solution, the photo-stationary *cis*-ratio in solid ($J/kT > 0$) shows a clear discontinuity (FIG. 2a) indicating a first-order phase transition. This is accompanied by the nucleation and growth as shown in FIG. 2b and Movie 2. Recall that the system starts with 0% *cis*-ratio. When the incoming light intensity is small, the molecular interaction suppresses almost all forward reaction by either photo activation or thermal relaxation because neighboring elements energetically favor the same state. Indeed, the lack of any initial *cis* creates a high energy barrier for nucleation to promote *trans*-to-*cis* isomerization. Once the light intensity is large enough to overcome this energy barrier, *cis*-nuclei form from a sufficient amount of *cis*-molecules. Subsequently, the nuclei keep growing with an auto-catalytic mechanism, until it reaches about 100% *cis*-ratio. The molecular interaction between *cis*-molecules further suppresses almost all backward reaction to *trans*.

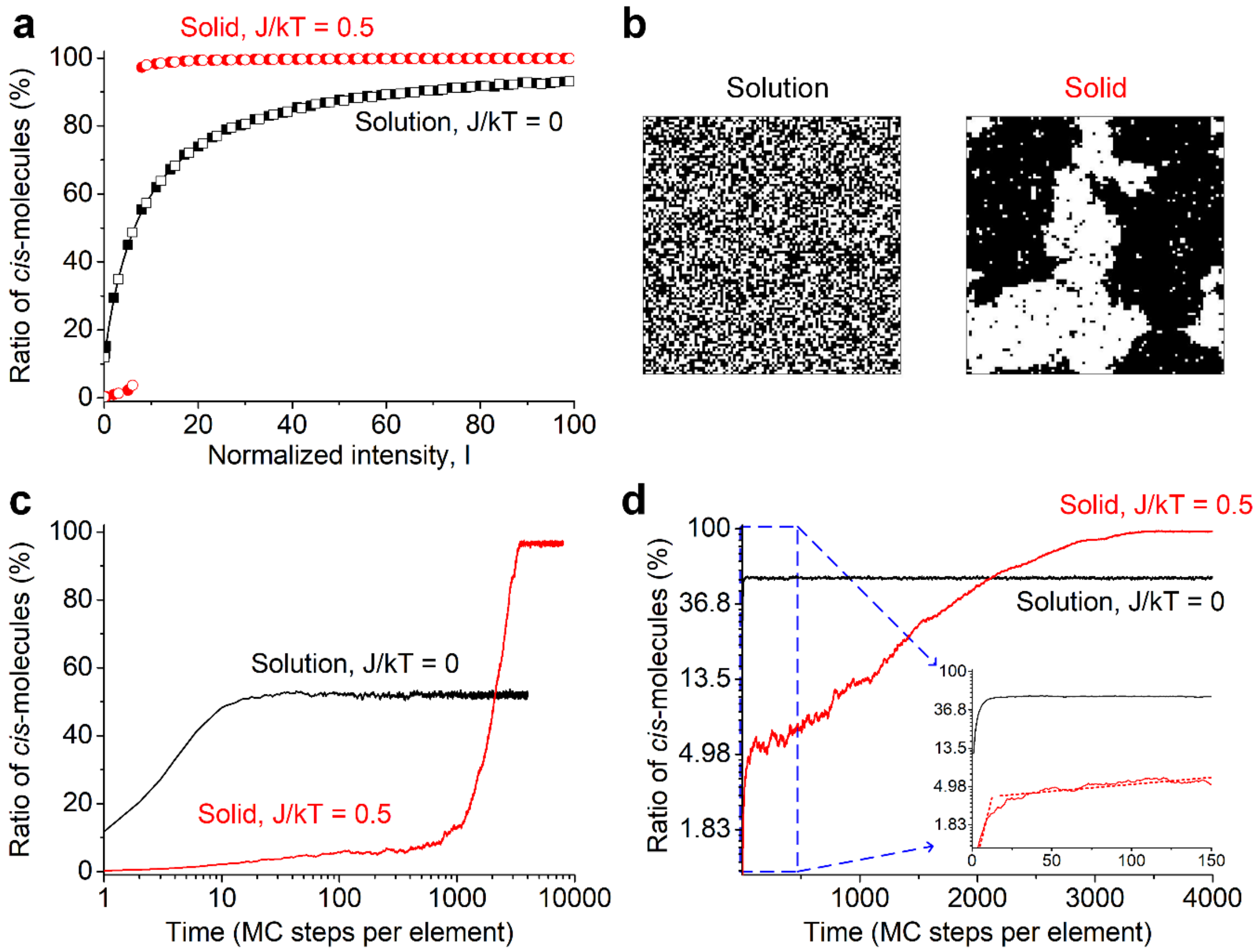

**FIG. 2.** Comparison between photoreactions in solution and solid, with fixed $h/kT = 1$. **(a)** The photo-stationary *cis*-ratio vs. light intensity. Filled circles/squares: original Hamiltonian (1). Open circles/squares: Photoreaction-incorporated Hamiltonian (3). Black solid line: analytical solution $\phi_{cis} = 1 - \left[1 + (I+1)\exp(-2h/kT)\right]^{-1}$. **(b)** Typical microstructure during photoreaction. Black square: *cis*. White square: *trans*. Also see Movies 1&2. **(c, d)** *cis*-ratio vs. time with fixed light intensity as $I = 7$ on logarithmic in time (**c**) and logarithmic in *cis*-ratio (**d**) scale.

We now show that the *trans*-to-*cis* transformation with increasing illumination in the solid is the manifestation of a first-order phase transition induced by external field. As noted in the case of solution, the single-molecule forward reaction rate is $(1-\phi_{cis})\exp(-2h/kT)(I+1)\tau^{-1}$ which may be rewritten as

$$(1-\phi_{cis})\exp\left(-\frac{2h - kT\log(I+1)}{kT}\right)\tau^{-1}. \tag{2}$$

Therefore, the light-driven forward reaction is equivalent to adding an additional external potential $-(1/2)kT\log(I+1)$ to the system. With this, the photoreaction-incorporated Hamiltonian is

$$H = -\frac{1}{2} J \sum_{\langle \alpha,\beta \rangle} S_\alpha S_\beta + \left(h - (1/2) kT \log(I+1)\right) \sum_{\alpha=1}^{n^2} S_\alpha + E\bar{W}(S_\alpha). \quad (3)$$

We conduct the same Monte Carlo simulation using this modified Hamiltonian with only thermal relaxation steps. The results (open dots in FIG. 2a) overlap perfectly with those from the original algorithm. Further, note that (3) is the classical Ising Hamiltonian when $E = 0$, and this undergoes a first-order phase transition induced by external field below the critical temperature which in our case is rescaled by illumination.

To further probe the effect of nucleation and growth, we compare the reaction kinetics. FIG. 2c and 2d plot the evolution using logarithmic scale on time in 2c and on *cis*-ratio in 2d. In solution, the *cis*-ratio grows following the first-order kinetics with a constant relaxation, reaching a finite photo-stationary plateau. In contrast, in solid, we observe initial fast exponential relaxation to a metastable state (inset of FIG. 2d), followed by a significantly slower sigmoidal kinetics of nucleation and growth. We attribute this slow kinetics to the lack of *cis*-nuclei initially in the system. Due to the single-flip nature of the simulation, forming an effective nucleus takes a long time. Once large enough *cis*-nuclei form, the system quickly goes through the auto-catalytic reaction towards the final photo-stationary state with nearly 100% *cis*-ratio. This sigmoidal kinetics in solid-state photoreaction has been experimentally observed recently in various material systems [24-26].

In addition to the sigmoidal kinetics, recent experiments on solid-state photo-dimerization also show that the reaction kinetics depends significantly on reaction history, while the photo-stationary state only depends on the final light intensity [26]. We are unaware of any existing model that is capable of capturing this path-dependent reaction kinetics. FIG. 3a shows the reaction kinetics (*trans*-ratio vs. time) under various light intensities and reaction histories. In one group of simulation (solid lines in FIG. 3a), we first apply a light with intensity $I = +\infty$ until reaching 5% *trans*-ratio, and then apply a weak light. In the other group (dashed lines in FIG. 3a), we directly apply the corresponding weak light. For each different weak light intensity, it is observed that both reaction paths (strong-weak vs. directly weak) lead to the same photo-stationary state. The photo-stationary *trans*-ratio only depends on the final weak light intensity. However, the reaction kinetics greatly depends on the reaction history. For instance, comparing the two curves at $I$ = 1/5, the photo-stationary state requires much longer time under a weak light alone compared to the strong-weak illumination. In the strong-weak illumination, one has both phases in coexistence and therefore does not need an initial nucleation event. Under weak illumination alone, the need for the initial nucleation leads to slower kinetics.

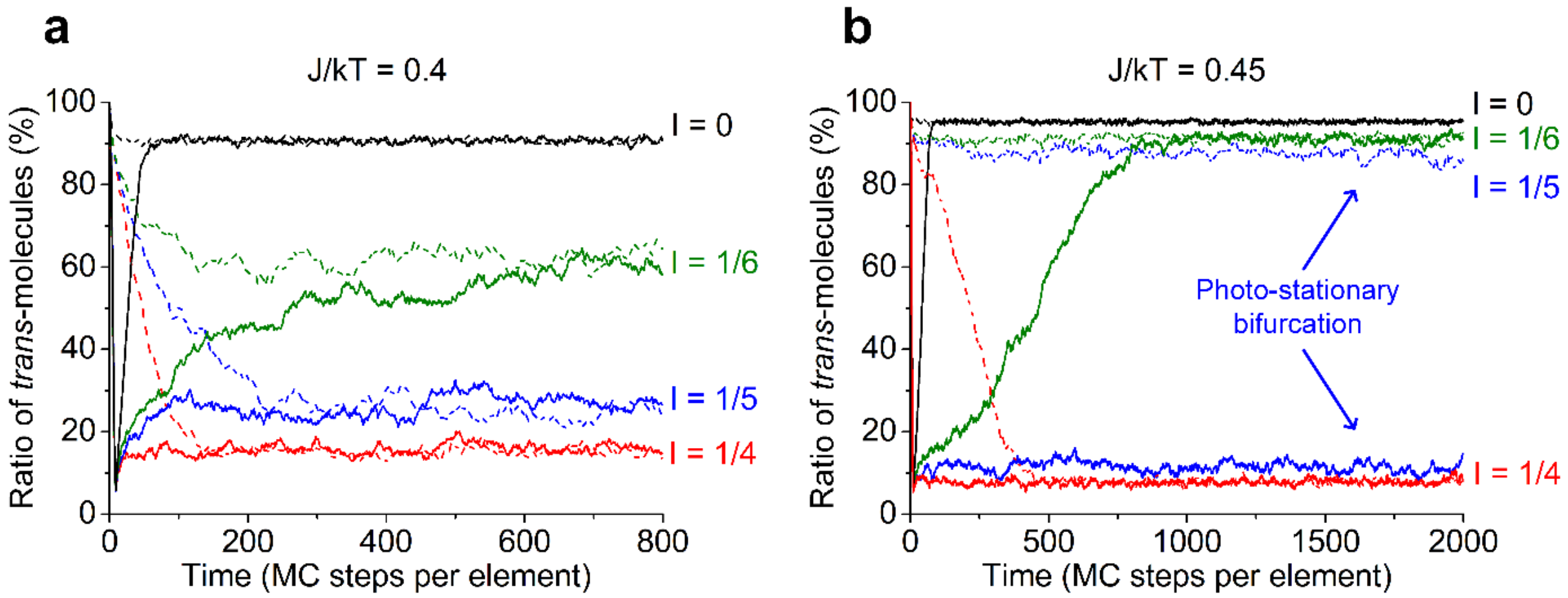

**FIG. 3.** Photoreaction kinetics under different reaction histories and intermolecular interactions $J/kT$. $h/J$ = 0.2 is fixed in both figures. **(a)** $J/kT$ = 0.4. **(b)** $J/kT$ = 0.45. All simulations start from 100% *trans*. Solid lines: strong-weak case where an initial strong illumination (of $I = +\infty$) is applied until we have 5% *trans*-ratio followed by a weak illumination. Dashed lines: weak case where only the weak illumination is applied. Note the bifurcation at $I$ =1/5 for the larger $J/kT$.

We then slightly increase $J/kT$ while keeping other conditions the same, and the results are shown in FIG. 3b. It is observed that all the photo-stationary states reach close to either 0% or 100% *trans*-ratio due to the stronger interaction. Further a bifurcation is observed for the intermediate illumination $I$ = 1/5 where the photo-stationary state is nearly all *trans* for strong-weak illumination but nearly all *cis* for weak illumination. This is again due to the stronger intermolecular interaction that suppresses nucleation.

We now turn to the case with elastic interactions ($E > 0$). Each element $\alpha$ located at $\mathbf{x}^\alpha$ is assigned with a photo-strain $\varepsilon_{ij}^{*\alpha}$. We choose two specific types of photo-strain, along with their corresponding dimensionless total elastic energies (assuming zero average stress) $\bar{W}\left(S_\alpha\right)$ as (see Supplementary Information for the detailed derivation)

Shear: 
$$\begin{cases} \varepsilon_{ij}^{*\alpha} = S_\alpha diag(1,-1), \\ \bar{W}(S_\alpha) = \dfrac{1}{2n^2}\sum_{\mathbf{k}\neq 0}\dfrac{(k_1^2-k_2^2)^2}{\mathbf{k}^4}\left|\tilde{S}(\mathbf{k})\right|^2. \end{cases} \tag{4}$$

Volumetric: 
$$\begin{cases} \varepsilon_{ij}^{*\alpha} = S_\alpha diag(1,1), \\ \bar{W}(S_\alpha) = \dfrac{n^2}{2}\left(\left\langle (S_\alpha)^2 \right\rangle - \langle S_\alpha \rangle^2\right). \end{cases} \tag{5}$$

$\tilde{S}(\mathbf{k})$ is the discrete Fourier transform of $S_\alpha$. The transformation strain in (4) for $S_\alpha = +1$ corresponds to an elongation in the $\boldsymbol{e}_1$ direction and compression in the $\boldsymbol{e}_2$ direction of the same magnitude. It can be viewed as a pure shear state if we rotate the coordinate system by $\pi/4$. The transformation strain in (5) corresponds to a 2D volumetric expansion or compression. The discrete Fourier transform $\tilde{S}(\mathbf{k})$ is

$$\tilde{S}(\mathbf{k}) = \sum_{\alpha=1}^{n^2} S_\alpha \exp\left(-i\mathbf{k}\cdot\mathbf{x}^\alpha\right), \tag{6}$$

where $\mathbf{k}$ is the coordinate in the Fourier space.

FIG. 4 shows the photo-stationary state vs. illumination for the two types of photo-strain with varying $E/kT$ and $J/kT$, together with their representative microstructures when $E/kT$ is large. When both $E/kT$ and $J/kT$ are small (e.g., $E/kT = J/kT = 0$), the photo-stationary state is similar to that shown in FIG. 2a. When $E/kT > 0$, the elastic energy is zero for the pure phases under zero average stress and this suppresses the mixed states. Note that the critical illumination for the light-induced *trans*-to-*cis* transition increases with increasing $E/kT$ and $J/kT$.

Crucially, by comparing FIG. 4a with 4c, and FIG. 4b with 4d, we see very significant differences in both the photo-stationary state and the morphology with the form of the photo-strain. First, when $J/kT = 0$ and the elastic interaction $E/kT$ is small ($\leq 2$), we see continuous transition in the case of shear (FIG. 4a) but discontinuous transition in the case of volumetric photo-strain (FIG. 4c). Second, while both cases show discontinuous transition at higher $E/kT$ or $J/kT$ increases, the critical illumination for the light-induced *trans*-to-*cis* transition is significantly higher in the case of volumetric photo-strain compared to the case of shear. This difference increases with increasing $E/kT$ and $J/kT$. Indeed, for the cases $J/kT = 0$ and $E/kT = 5$ (FIG. 4a and 4c) and $J/kT = 0.4$ and $E/kT = 2$ (FIG. 4b and 4d), we see light-induced transition in the case of shear photo-strain, but no light-induced transition in the case of volumetric photo-strain in the range of illumination considered. Therefore, *the nature of light-induced behavior can depend critically on the exact form of the photo-strain.*

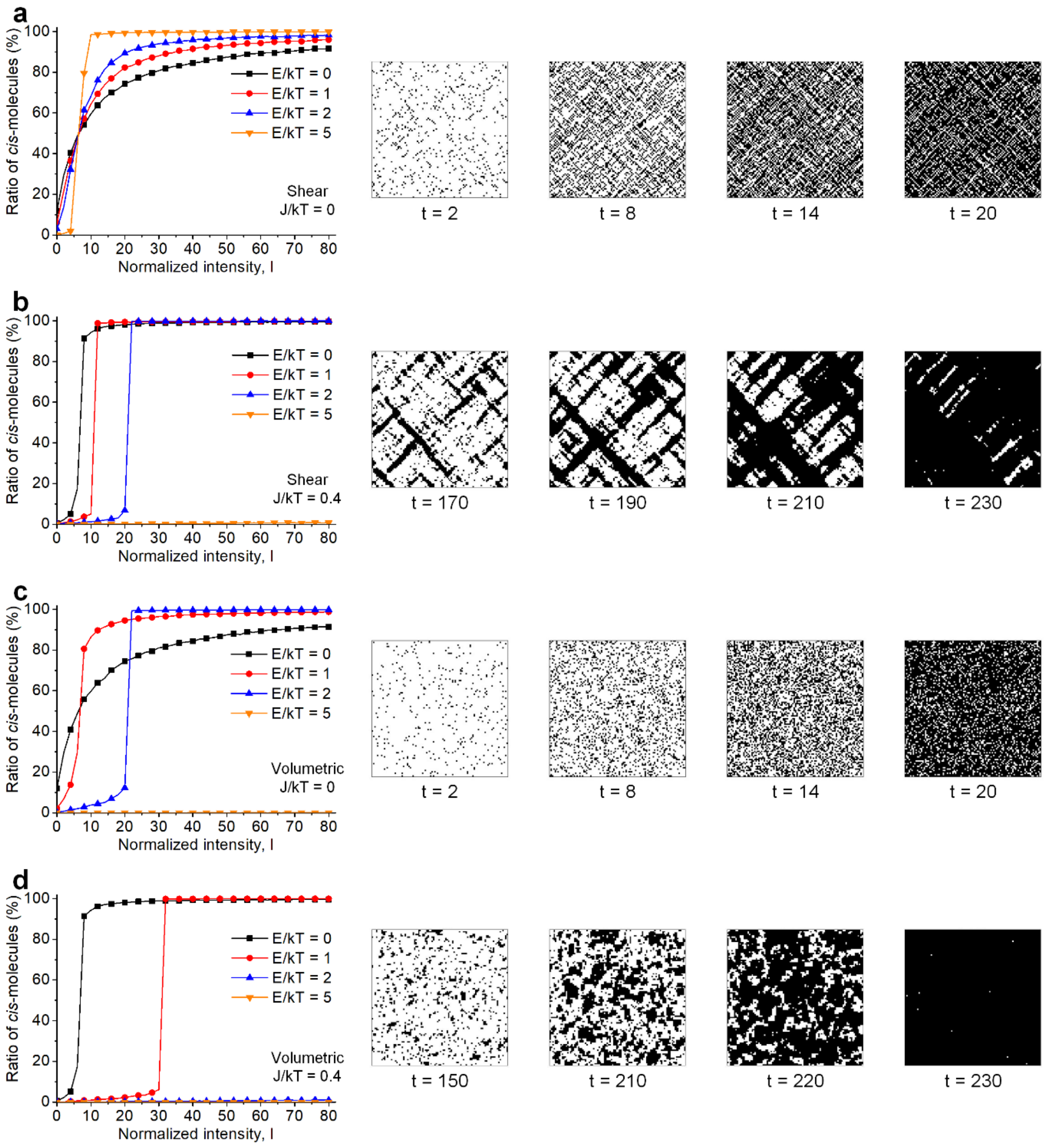

**FIG. 4.** Photo-stationary states considering long-range elastic interaction with different $E/kT$ and $J/kT$, where $h/kT = 1$ is fixed. **(a)** Shear photo-strain, $J/kT = 0$. **(b)** Shear photo-strain, $J/kT = 0.4$. Also see Movie 3. **(c)** Volumetric photo-strain, $J/kT = 0$. **(d)** Volumetric photo-strain, $J/kT = 0.4$. Also see Movie 4. On the right of each diagram shows the representative kinetic patterns during the reaction, with $E/kT = 2$ in shear, and $E/kT = 1$ in volumetric. The time $t$ represents the accumulated MC steps per element. Each black or white square represents a *cis*- or *trans*-element, respectively.

We also see significant differences in the morphology during the light-induced transition in the two cases. In the case of shear photo-strain (FIG. 4a and 4b, Movie 3), we find a "tweed" microstructure of crossing diagonal stripes while in the case of volumetric photo-strain (FIG. 4c and 4d, Movie 4), we find a more disorganized pattern. In both cases, the features are very fine (at the scale of pixels) when the nearest-neighbor interaction $J/kT = 0$ (FIG. 4a and 4c), but takes on a finite scale with non-zero $J/kT$ (FIG. 4b and 4d). Further the morphology is different in the presence of elasticity compared to that in its absence (compare FIG. 4 with FIG. 2b).

To understand the differences between the two cases of photo-strain, recall the Hadamard jump condition [38, 39]. According to this condition two domains with strain $\boldsymbol{\varepsilon}^+$ and $\boldsymbol{\varepsilon}^-$ can coexist across a boundary with normal $\mathbf{n}$ if there exists a vector $\mathbf{a}$ such that

$$\boldsymbol{\varepsilon}^+ - \boldsymbol{\varepsilon}^- = \frac{1}{2}\left(\mathbf{a} \otimes \mathbf{n} + \mathbf{n} \otimes \mathbf{a}\right). \tag{7}$$

With shear photo-strain, neighboring domains with different shear can coexist with zero stress if the boundary between them has a normal $\mathbf{n} = (\pm 1, 1)^T$. Therefore, the formation of stripe or tweed domains can ameliorate the elastic energy as in first-order phase transition in martensites [40-42]. This allows for first-order kinetics for small $E/kT$ and small $J/kT$ and a smaller critical illumination at large $E/kT$. In contrast, with volumetric photo-strain, there are no vectors $\mathbf{a}$ and $\mathbf{n}$ that enable us to solve (7). Therefore, the elastic energy cannot be ameliorated and this impedes nucleation when $E/kT > 0$. In both cases, the nearest-neighbor interaction $J$ acts as the effective interfacial energy between neighboring domains. A larger $J/kT$ thus induces a larger length scale of the initial domain size, leading to the coarsening of domains. This collective behavior of domain pattern formation in solids has been observed experimentally in azobenzene self-assembly monolayers [28, 29]. The results here are also consistent with the observation in molecular crystals of photo-active molecules: some so-called photo-salient crystals shatter when illuminated [43] while others undergo photo-actuation without shattering [44]. The difference has been shown to be consistent with the ability of the photo-strain to satisfy the Hadamard jump condition [45].

In addition to the long-range elastic interaction, many photoactive molecules interact under the long-range electrostatic dipole-dipole interaction. For example, an azobenzene isomer can be modified with additional functional groups to carry a permanent dipole moment [29]. The energy due to the electrostatic dipole-dipole interaction can be calculated following the same methodology proposed in our current model. In that case, the domain formation is expected to depend additionally on the distribution of dipole moments.

In summary, we have postulated an Ising-like lattice with Monte Carlo simulation to explore the collective behavior in solid-state photoreaction. Unlike photoreactions in solution following a first-order kinetics, photoreaction in solid shows a feature analogous to first-order phase transition: a sigmoidal kinetics of nucleation and growth. This leads to a bifurcation in the photo-stationary state. The photo-strain upon reaction significantly affects the kinetics, equilibrium, and formation of microstructure through the long-range elastic interaction. These predictions agree qualitatively well with recent experimental results. This study is hoped to help the future design and optimization of solid-based photomechanical systems.

**Acknowledgement**

We gratefully acknowledge many stimulating discussions with Christopher Bardeen. This work was supported by the Office of Naval Research through the MURI on Photomechanical Material Systems (ONR N00014-18-1-2624).

## Supplementary Information

## Collective behavior in the kinetics and equilibrium of solid-state photoreaction

Ruobing Bai, Ying Shi Teh, Kaushik Bhattacharya*

Division of Engineering and Applied Science, California Institute of Technology, Pasadena, California 91125, USA
* Corresponding author: bhatta@caltech.edu

**Movie 1.** Representative microstructure evolution in solution with $J/kT = 0$, $E/kT = 0$, $h/kT = 1$.

**Movie 2.** Representative microstructure evolution in solid with $J/kT = 0.5$, $E/kT = 0$, $h/kT = 1$.

**Movie 3.** Representative microstructure evolution with the shear photo-strain, $J/kT = 0.4$, $E/kT = 2$, and $h/kT = 1$.

**Movie 4.** Representative microstructure evolution with the volumetric photo-strain, $J/kT = 0.4$, $E/kT = 1$, and $h/kT = 1$.

### 1. Fourier series and Fourier transform

We consider a periodic square cell of length $L$ and volume $V$ ($d$-dimension) extended infinitely. A periodic function $f(\mathbf{x})$ associated with it can be expressed by the Fourier series:

$$f(\mathbf{x}) = \frac{1}{V}\sum_{\mathbf{k}} \tilde{f}(\mathbf{k})\exp(i\mathbf{k}\cdot\mathbf{x}), \tag{S1}$$

where $\mathbf{k} = (2\pi / L)\mathbf{l}$ and $\mathbf{l}$ represents any lattice vector in the Fourier space. The Fourier transform of $f(\mathbf{x})$ is

$$\tilde{f}(\mathbf{k}) = \int_V f(\mathbf{x})\exp(-i\mathbf{k}\cdot\mathbf{x})\,d\mathbf{x}. \tag{S2}$$

### 2. Elastic energy due to a field of transformation strain

Now we consider an infinite linear elastic body with a periodic cell of volume $V$. The body undergoes a field of transformation strain (e.g., photo-strain induced by photoreaction), described by $\varepsilon_{ij}^{*}(\mathbf{x})$. The transformation strain is defined as the spontaneous strain of the material under no external stress. We assume $\varepsilon_{ij}^{*}(\mathbf{x})$ to be periodic with $V$.

We further assume that before and after the transformation strain, the stiffness tensor in the material,

$C_{ijkl}$, remains unchanged and homogeneous. From Eshelby et al. [1,2], the current problem is equivalent to an elastic body under a distribution of body force

$$P_i(\mathbf{x}) = -\frac{\partial C_{ijkl}\varepsilon_{kl}^*(\mathbf{x})}{\partial x_j}. \tag{S3}$$

The strain field $\varepsilon_{ij}(\mathbf{x})$ is related to the displacement field $u_i(\mathbf{x})$ via

$$\varepsilon_{ij}(\mathbf{x}) = \frac{1}{2}\left(\frac{\partial u_i}{\partial x_j} + \frac{\partial u_j}{\partial x_i}\right), \tag{S4}$$

and the stress field is

$$\sigma_{ij}(\mathbf{x}) = C_{ijkl}\left(\varepsilon_{kl} - \varepsilon_{kl}^*\right). \tag{S5}$$

The balance of force written using the displacement field is

$$C_{ijkl}\frac{\partial^2 u_k(\mathbf{x})}{\partial x_j \partial x_l} = C_{ijkl}\frac{\partial \varepsilon_{kl}^*(\mathbf{x})}{\partial x_j} = \frac{\partial M_{ij}(\mathbf{x})}{\partial x_j}, \tag{S6}$$

where $M_{ij}(\mathbf{x}) = C_{ijkl}\varepsilon_{kl}^*(\mathbf{x})$ is defined as the elastic dipole moment induced by $\varepsilon_{ij}^*(\mathbf{x})$. The elastic energy stored in the periodic cell $V$ is

$$W = \frac{1}{2}\int_V C_{ijkl}\left(\varepsilon_{ij} - \varepsilon_{ij}^*\right)\left(\varepsilon_{kl} - \varepsilon_{kl}^*\right) d\mathbf{x}. \tag{S7}$$

Because $\varepsilon_{ij}^*(\mathbf{x})$ is periodic, the induced strain field $\varepsilon_{ij}(\mathbf{x})$ is also periodic. As a result, we split $\varepsilon_{ij}(\mathbf{x})$ into two parts, $\varepsilon_{ij}(\mathbf{x}) = \hat{\varepsilon}_{ij}(\mathbf{x}) + \langle \varepsilon_{ij} \rangle$, satisfying

$$\int_V \hat{\varepsilon}_{ij}(\mathbf{x}) d\mathbf{x} = 0, \tag{S8}$$

$$\frac{1}{V}\int_V \varepsilon_{ij}(\mathbf{x}) d\mathbf{x} = \langle \varepsilon_{ij} \rangle. \tag{S9}$$

The displacement field $\hat{u}_i(\mathbf{x})$ corresponding to $\hat{\varepsilon}_{ij}(\mathbf{x})$ is thereby periodic, while the displacement field corresponding to $\langle \varepsilon_{ij} \rangle$ is not periodic. With this, we rewrite the elastic energy

$$
\begin{aligned}
W &= \frac{1}{2}\int_V C_{ijkl}\left(\varepsilon_{ij}-\varepsilon_{ij}^*\right)\left(\varepsilon_{kl}-\varepsilon_{kl}^*\right)d\mathbf{x} \\
&= \hat{W} + \frac{V}{2}C_{ijkl}\left\langle\varepsilon_{ij}\right\rangle\left\langle\varepsilon_{kl}\right\rangle - VC_{ijkl}\left\langle\varepsilon_{ij}\right\rangle\left\langle\varepsilon_{kl}^*\right\rangle,
\end{aligned}
\tag{S10}
$$

where $\hat{W} = \frac{1}{2}\int_V C_{ijkl}\left(\hat{\varepsilon}_{ij}-\varepsilon_{ij}^*\right)\left(\hat{\varepsilon}_{kl}-\varepsilon_{kl}^*\right)d\mathbf{x}$ is the elastic energy associated with $\hat{\varepsilon}_{ij}\left(\mathbf{x}\right)$, and the rest only depend on the average terms $\left\langle\varepsilon_{ij}\right\rangle$ and $\left\langle\varepsilon_{ij}^*\right\rangle$.

**2.1 Elastic energy related to the average strain**

In the current problem, we prescribe a far field condition that the average stress in the periodic cell is zero

$$
\left\langle\sigma_{ij}\right\rangle = C_{ijkl}\left\langle\varepsilon_{kl}-\varepsilon_{kl}^*\right\rangle = 0\,. \tag{S11}
$$

Noticing $\varepsilon_{ij}\left(\mathbf{x}\right) = \hat{\varepsilon}_{ij}\left(\mathbf{x}\right) + \left\langle\varepsilon_{ij}\right\rangle$ and $\left\langle\hat{\varepsilon}_{ij}\right\rangle = 0$, from (S11) we get $\left\langle\varepsilon_{ij}\right\rangle = \left\langle\varepsilon_{ij}^*\right\rangle$. As a result, the last two terms in (S10) is simplified to be

$$
\frac{V}{2}C_{ijkl}\left\langle\varepsilon_{ij}\right\rangle\left\langle\varepsilon_{kl}\right\rangle - VC_{ijkl}\left\langle\varepsilon_{ij}\right\rangle\left\langle\varepsilon_{kl}^*\right\rangle = -\frac{V}{2}C_{ijkl}\left\langle\varepsilon_{ij}^*\right\rangle\left\langle\varepsilon_{kl}^*\right\rangle. \tag{S12}
$$

**2.2 Elastic energy $\hat{W}$** associated with periodic displacement

To further calculate $\hat{W}$ in (S10), we rewrite it as $\hat{W} = W_a + W_b + W_c$, where

$$
\begin{cases}
W_a = \frac{1}{2}\int_V C_{ijkl}\hat{\varepsilon}_{ij}\hat{\varepsilon}_{kl}d\mathbf{x}, \\
W_b = -\int_V C_{ijkl}\varepsilon_{ij}^*\hat{\varepsilon}_{kl}d\mathbf{x}, \\
W_c = \frac{1}{2}\int_V C_{ijkl}\varepsilon_{ij}^*\varepsilon_{kl}^*d\mathbf{x}.
\end{cases}
\tag{S13}
$$

The balance of force in (S6) can be expressed using only the periodic displacement $\hat{u}_i\left(\mathbf{x}\right)$

$$
C_{ijkl}\frac{\partial^2\hat{u}_k}{\partial x_j\partial x_l} = C_{ijkl}\frac{\partial\varepsilon_{kl}^*}{\partial x_j}\,. \tag{S14}
$$

With this, and applying the divergence theorem and the periodicity of $\hat{\varepsilon}_{ij}$ and $\hat{u}_i$, we rewrite $W_a$ as

$$W_a = \frac{1}{2}\int_V C_{ijkl}\varepsilon_{ij}^*\hat{\varepsilon}_{kl}d\mathbf{x}, \tag{S15}$$

The elastic energy $\hat{W}$ is therefore

$$\hat{W} = \frac{1}{2}\int_V C_{ijkl}\varepsilon_{ij}^*\varepsilon_{kl}^*d\mathbf{x} - \frac{1}{2}\int_V C_{ijkl}\varepsilon_{ij}^*\hat{\varepsilon}_{kl}d\mathbf{x} = W_0 + W_1, \tag{S16}$$

where we have denoted

$$\begin{cases} W_0 = \frac{1}{2}\int_V C_{ijkl}\varepsilon_{ij}^*\varepsilon_{kl}^*d\mathbf{x}, \\ W_1 = -\frac{1}{2}\int_V C_{ijkl}\varepsilon_{ij}^*\hat{\varepsilon}_{kl}d\mathbf{x}. \end{cases} \tag{S17}$$

The term $W_0$ contains only the "self-interaction" elastic energy rising from $\varepsilon_{ij}^*$, while the term $W_1$ contains the energy of elastic interaction between transformation strains at different locations. Hence, we need to further calculate $W_1$.

Because $\hat{\varepsilon}_{ij}(\mathbf{x})$ is periodic, we can expand it using the Fourier series

$$\hat{\varepsilon}_{ij}(\mathbf{x}) = \frac{1}{V}\sum_{\mathbf{k}}\tilde{\hat{\varepsilon}}_{ij}(\mathbf{k})\exp(i\mathbf{k}\cdot\mathbf{x}), \tag{S18}$$

with $\mathbf{k} = (2\pi / L)\mathbf{l}$, $\mathbf{l}$ as the lattice vectors in the Fourier space, and

$$\tilde{\hat{\varepsilon}}_{ij}(\mathbf{k}) = \int_V \hat{\varepsilon}_{ij}(\mathbf{x})\exp(-i\mathbf{k}\cdot\mathbf{x})d\mathbf{x}. \tag{S19}$$

The summation in (S18) is summed up to infinity, but within the finite Fourier domain when approximated using the discrete Fourier transform numerically. From (S19), we have $\tilde{\hat{\varepsilon}}_{ij}(\mathbf{k}=0) = \int_V \hat{\varepsilon}_{ij}(\mathbf{x})d\mathbf{x} = 0$, so we can express

$$\hat{\varepsilon}_{ij}(\mathbf{x}) = \frac{1}{V}\sum_{\mathbf{k}\neq 0}\tilde{\hat{\varepsilon}}_{ij}(\mathbf{k})\exp(i\mathbf{k}\cdot\mathbf{x}). \tag{S20}$$

Substituting this into $W_1$, we obtain

$$W_1 = -\frac{1}{2V}\sum_{\mathbf{k}\neq 0}C_{ijkl}\int_V \varepsilon_{ij}^*(\mathbf{x})\tilde{\hat{\varepsilon}}_{kl}(\mathbf{k})\exp(i\mathbf{k}\cdot\mathbf{x})d\mathbf{x}. \tag{S21}$$

Notice that the Fourier transform of $\varepsilon_{ij}^{*}(\mathbf{x})$ is

$$\tilde{\varepsilon}_{ij}^{*}(\mathbf{k})=\int_{V}\varepsilon_{ij}^{*}(\mathbf{x})\exp(-i\mathbf{k}\cdot\mathbf{x})d\mathbf{x}. \tag{S22}$$

We can further express $W_1$ as

$$W_1=-\frac{1}{2V}\sum_{\mathbf{k}\neq 0}C_{ijkl}\bar{\tilde{\varepsilon}}_{kl}^{*}(\mathbf{k})\tilde{\hat{\varepsilon}}_{ij}(\mathbf{k}), \tag{S23}$$

where $\bar{\tilde{\varepsilon}}_{kl}^{*}(\mathbf{k})$ is the conjugate of $\tilde{\varepsilon}_{kl}^{*}(\mathbf{k})$.

Because $\varepsilon_{ij}^{*}(\mathbf{x})$ is a prescribed field in this problem, its Fourier transform $\tilde{\varepsilon}_{ij}^{*}(\mathbf{k})$ is also known. Therefore, we only need to solve for $\tilde{\hat{\varepsilon}}_{ij}(\mathbf{k})$. Using the property of Fourier transform and the symmetry of $C_{ijkl}$, we can replace $\tilde{\hat{\varepsilon}}_{ij}(\mathbf{k})$ by $(ik_j)\tilde{\hat{u}}_i(\mathbf{k})$ when it is multiplied by $C_{ijkl}$. Rewriting the balance of force in (S14) in the Fourier space, and recalling that $M_{ij}(\mathbf{x})=C_{ijkl}\varepsilon_{kl}^{*}(\mathbf{x})$, we have

$$C_{ijkl}(ik_j)(ik_l)\tilde{\hat{u}}_k=(ik_j)\tilde{M}_{ij}, \tag{S24}$$

or

$$C_{ijkl}k_jk_l\tilde{\hat{u}}_k=-ik_j\tilde{M}_{ij}. \tag{S25}$$

Since $C_{ijkl}k_jk_l$ is positive definite and symmetric for any $\mathbf{k}\neq 0$, we can solve for $\tilde{\hat{u}}_k$ using its inverse. For a homogeneous isotropic linear elastic material,

$$C_{ijkl}=\lambda\delta_{ij}\delta_{kl}+\mu(\delta_{ik}\delta_{jl}+\delta_{il}\delta_{jk}), \tag{S26}$$

where the Lame constant $\lambda$ follows the relation

$$\lambda=\frac{2\mu\nu}{1-2\nu}, \tag{S27}$$

and $\mu$ and $\nu$ are the shear modulus and Poisson's ratio. With these, we obtain

$$\tilde{\hat{u}}_i(\mathbf{k})=\frac{-i}{\mu}\tilde{M}_{kl}\left[\frac{k_k}{\mathbf{k}^2}\delta_{il}-\frac{1}{2(1-\nu)}\frac{k_ik_kk_l}{\mathbf{k}^4}\right], \text{ for } \mathbf{k}\neq 0. \tag{S28}$$

Substituting (S28) into $\tilde{\tilde{\varepsilon}}_{ij}(\mathbf{k}) = (ik_j)\tilde{\tilde{u}}_i(\mathbf{k})$, and further into $W_1$ in (S23), we obtain

$$\begin{aligned} W_1 &= -\frac{1}{2V}\sum_{\mathbf{k}\neq 0} C_{ijkl}\bar{\tilde{\varepsilon}}_{kl}^*(\mathbf{k})\tilde{\tilde{\varepsilon}}_{ij}(\mathbf{k}) \\ &= -\frac{1}{2\mu V}\sum_{\mathbf{k}\neq 0}\bar{\tilde{M}}_{ij}(\mathbf{k})\tilde{M}_{kl}(\mathbf{k})\left[\frac{k_j k_k}{\mathbf{k}^2}\delta_{il} - \frac{1}{2(1-\nu)}\frac{k_i k_j k_k k_l}{\mathbf{k}^4}\right]. \end{aligned} \tag{S29}$$

To sum up, the total elastic energy under the far field condition of zero average stress is

$$W = \frac{1}{2}\int_V C_{ijkl}\varepsilon_{ij}^*\varepsilon_{kl}^* d\mathbf{x} + W_1 - \frac{V}{2}C_{ijkl}\left\langle\varepsilon_{ij}^*\right\rangle\left\langle\varepsilon_{kl}^*\right\rangle, \tag{S30}$$

with

$$W_1 = -\frac{1}{2\mu V}\sum_{\mathbf{k}\neq 0}\bar{\tilde{M}}_{ij}(\mathbf{k})\tilde{M}_{kl}(\mathbf{k})\left[\frac{k_j k_k}{k^2}\delta_{il} - \frac{1}{2(1-\nu)}\frac{k_i k_j k_k k_l}{k^4}\right]. \tag{S31}$$

The total elastic energy contains three parts, the purely self-interacting energy, the energy containing the non-trivial elastic interaction, and the energy rising from the far-field condition (zero average stress). When considering a far-field condition of zero average strain, one can show that the third term in (S30) vanishes.

**3.1 Elastic interaction due to shear transformation strain**

Now we consider the 2D lattice in the letter with a field of transformation strain

$$\varepsilon_{ij}^*(\mathbf{x}) = \begin{bmatrix} S(\mathbf{x}) & \\ & -S(\mathbf{x}) \end{bmatrix}. \tag{S32}$$

This is equivalent to a strain state of pure shear if we rotate the coordinate system by $\pi/4$. By rescaling, we let the magnitude of $S(\mathbf{x})$ at $\mathbf{x} = \mathbf{x}^\alpha$ be the value of the element $S_\alpha$, where $\mathbf{x}^\alpha$ is the coordinate of element $\alpha$.

The elastic dipole moment in this case is

$$M_{ij}(\mathbf{x}) = C_{ijkl}\varepsilon_{kl}^*(\mathbf{x}) = 2\mu\begin{bmatrix} S(\mathbf{x}) & \\ & -S(\mathbf{x}) \end{bmatrix} = 2\mu\left(\delta_{i1}\delta_{j1} - \delta_{i2}\delta_{j2}\right)S(\mathbf{x}). \tag{S33}$$

Correspondingly, we can define the scalar transformation strain and elastic dipole moment in the Fourier space as $\tilde{S}(\mathbf{k})$ and $\tilde{M}_{ij}(\mathbf{k}) = 2\mu\left(\delta_{i1}\delta_{j1} - \delta_{i2}\delta_{j2}\right)\tilde{S}(\mathbf{k})$. Substituting this into (S31), we get

$$W_1 = -\frac{2\mu}{V}\sum_{\mathbf{k}\neq 0}\left[1-\frac{1}{2(1-\nu)}\frac{\left(k_1^2-k_2^2\right)^2}{\mathbf{k}^4}\right]\left|\tilde{S}(\mathbf{k})\right|^2 . \tag{S34}$$

Further note that in this case

$$\frac{1}{2}\int_V C_{ijkl}\varepsilon_{ij}^*\varepsilon_{kl}^* d\mathbf{x} = 2\mu\int_V S(\mathbf{x})^2 d\mathbf{x} = 2\mu V\left\langle S^2\right\rangle , \tag{S35}$$

and

$$\frac{V}{2}C_{ijkl}\left\langle \varepsilon_{ij}^*\right\rangle\left\langle \varepsilon_{kl}^*\right\rangle = 2\mu V\left\langle S\right\rangle^2 . \tag{S36}$$

Substituting these into (S30), we obtain the total energy

$$\begin{aligned} W &= \frac{1}{2}\int_V C_{ijkl}\varepsilon_{ij}^*\varepsilon_{kl}^* d\mathbf{x} + W_1 - \frac{V}{2}C_{ijkl}\left\langle \varepsilon_{ij}^*\right\rangle\left\langle \varepsilon_{kl}^*\right\rangle \\ &= 2\mu V\left\langle S^2\right\rangle + W_1 - 2\mu V\left\langle S\right\rangle^2 . \end{aligned} \tag{S37}$$

We can then use the relation

$$\sum_{\mathbf{k}\neq 0}\left|\tilde{S}(\mathbf{k})\right|^2 = \sum_{\mathbf{k}}\left|\tilde{S}(\mathbf{k})\right|^2 - \left|\tilde{S}(\mathbf{k}=0)\right|^2 = V^2\left\langle S^2\right\rangle - V^2\left\langle S\right\rangle^2 . \tag{S38}$$

Substituting this into (S34), we get

$$W_1 = -2\mu V\left\langle S^2\right\rangle + 2\mu V\left\langle S\right\rangle^2 + \frac{\mu}{V(1-\nu)}\sum_{\mathbf{k}\neq 0}\frac{\left(k_1^2-k_2^2\right)^2}{\mathbf{k}^4}\left|\tilde{S}(\mathbf{k})\right|^2 . \tag{S39}$$

The total elastic energy in (S37) is further simplified as

$$W = \frac{\mu}{V(1-\nu)}\sum_{\mathbf{k}\neq 0}\frac{\left(k_1^2-k_2^2\right)^2}{\mathbf{k}^4}\left|\tilde{S}(\mathbf{k})\right|^2 . \tag{S40}$$

In the simulation, the approximated discrete Fourier transform $\tilde{S}(\mathbf{k})$ is

$$\tilde{S}(\mathbf{k}) = \sum_{\alpha=1}^{n^2} S_\alpha \exp\left(-i\mathbf{k}\cdot\mathbf{x}^\alpha\right), \tag{S41}$$

where $S_\alpha$ is the value of element $\alpha$.

We finally replace the coefficient $\mu/\left[V(1-\nu)\right]$ with $E/\left(2n^2\right)$, where $E = 2\mu/(1-\nu)$ is the plane strain Young's modulus of the material, and the dimensionless volume of the periodic system is $V = n^2$. The total elastic energy from (S40) becomes

$$W = \frac{E}{2n^2}\sum_{\mathbf{k}\neq 0}\frac{\left(k_1^2-k_2^2\right)^2}{\mathbf{k}^4}\left|\tilde{S}(\mathbf{k})\right|^2 . \tag{S42}$$

When considering a far-field condition of zero average strain, by dropping the third term in (S30) we get

$$W = E(1-\nu)n^2\left\langle \varepsilon^*\right\rangle^2 + \frac{E}{2n^2}\sum_{\mathbf{k}\neq 0}\frac{\left(k_1^2-k_2^2\right)^2}{\mathbf{k}^4}\left|\tilde{S}(\mathbf{k})\right|^2 . \tag{S43}$$

**3.2 Elastic interaction due to volumetric transformation strain**

We next consider the case of volumetric transformation strain in 2D

$$\varepsilon_{ij}^*(\mathbf{x}) = S(\mathbf{x})\delta_{ij} . \tag{S44}$$

The elastic dipole moment is

$$M_{ij}(\mathbf{x}) = C_{ijkl}\varepsilon_{kl}^*(\mathbf{x}) = \frac{2\mu}{(1-2\nu)}S(\mathbf{x})\delta_{ij} . \tag{S45}$$

Correspondingly, we can define the scalar transformation strain and elastic dipole moment in the Fourier space as $\tilde{S}(\mathbf{k})$ and $\tilde{M}_{ij}(\mathbf{k})$ as in Section S3.1. Substituting this into (S31), we get

$$W_1 = -\frac{1}{V}\frac{\mu}{(1-\nu)(1-2\nu)}\sum_{\mathbf{k}\neq 0}\left|\tilde{S}(\mathbf{k})\right|^2 . \tag{S46}$$

We now include the term corresponding to $\mathbf{k} = 0$, such that

$$W_1 = -\frac{1}{V}\frac{\mu}{(1-\nu)(1-2\nu)}\sum_{\mathbf{k}}\left|\tilde{S}(\mathbf{k})\right|^2 + \frac{1}{V}\frac{\mu}{(1-\nu)(1-2\nu)}\left|\tilde{S}(\mathbf{k}=0)\right|^2 . \tag{S47}$$

Again, note that

$$\tilde{S}(\mathbf{k}=0) = \int_V S(\mathbf{x})d\mathbf{x} = V\left\langle S\right\rangle , \tag{S48}$$

so the second term becomes

$$\frac{\mu}{(1-\nu)(1-2\nu)} V \langle S \rangle^2 . \tag{S49}$$

For the first term, notice that

$$S(\mathbf{x}) = \frac{1}{V} \sum_{\mathbf{k}} \tilde{S}(\mathbf{k}) \exp(i\mathbf{k}\cdot\mathbf{x}), \tag{S50}$$

and

$$\int_V [S(\mathbf{x})]^2 dV = \frac{1}{V^2} \int_V \left| \sum_{\mathbf{k}} \tilde{S}(\mathbf{k}) \exp(i\mathbf{k}\cdot\mathbf{x}) \right|^2 dV = \frac{1}{V} \sum_{\mathbf{k}} \left| \tilde{S}(\mathbf{k}) \right|^2, \tag{S51}$$

where we have used the orthogonality of the Fourier series. Substituting (S48) and (S51) into (S47), we have

$$W_1 = \frac{\mu}{(1-\nu)(1-2\nu)} V \left( \langle S \rangle^2 - \langle S^2 \rangle \right), \tag{S52}$$

where we have expressed

$$\int_V [S(\mathbf{x})]^2 dV = V \langle S^2 \rangle . \tag{S53}$$

From (S30), the total elastic energy is

$$\begin{aligned} W &= \frac{1}{2} \int_V C_{ijkl} \varepsilon_{ij}^* \varepsilon_{kl}^* d\mathbf{x} + W_1 - \frac{V}{2} C_{ijkl} \langle \varepsilon_{ij}^* \rangle \langle \varepsilon_{kl}^* \rangle \\ &= \frac{\mu}{(1-\nu)} V \left( \langle S^2 \rangle - \langle S \rangle^2 \right). \end{aligned} \tag{S54}$$

Again, replacing $\mu/(1-\nu)$ with $E/2$ and $V = n^2$, we obtain

$$W = \frac{E}{2} n^2 \left( \langle S^2 \rangle - \langle S \rangle^2 \right). \tag{S55}$$

Expressing (S55) using the element value $S_\alpha$ leads to

$$W = \frac{E}{2} n^2 \left( \langle (S_\alpha)^2 \rangle - \langle S_\alpha \rangle^2 \right). \tag{S56}$$

When considering a far-field condition of zero average strain, by dropping the third term in (S30) we get

$$W = \frac{E}{2} n^2 \left( \frac{\langle S_\alpha \rangle^2}{(1-2\nu)} + \langle (S_\alpha)^2 \rangle \right). \tag{S57}$$

### 3.3 Computation of elastic energy

We can define a "stress" field to compute the total elastic energy and its change during the Monte Carlo element flip. Take the case of shear transformation strain in 3.1 as an example. From (S42), the total energy is expressed as

$$W = \frac{E}{2n^2} \sum_{\mathbf{k} \neq 0} \frac{\left(k_1^2 - k_2^2\right)^2}{\mathbf{k}^4} \left| \tilde{S}(\mathbf{k}) \right|^2 . \tag{S58}$$

We define a "stress" field as

$$\sigma_\alpha = \frac{\partial W}{\partial S_\alpha} = \frac{E}{n^2} \sum_{\mathbf{k} \neq 0} \frac{\left(k_1^2 - k_2^2\right)^2}{\mathbf{k}^4} \mathrm{Re}\left( \tilde{S}(\mathbf{k}) \exp\left( i\mathbf{k} \cdot \mathbf{x}^\alpha \right) \right). \tag{S59}$$

One can further show that the total elastic energy is calculated as

$$W = \frac{1}{2} \sum_{\alpha=1}^{n^2} \sigma_\alpha S_\alpha . \tag{S60}$$